# SVOM pointing strategy: how to optimize the redshift measurements?

B. Cordier*, F. Desclaux†, J. Foliard† and S. Schanne*

*CEA Saclay, DSM / IRFU / SAp, 91191 Gif sur Yvette, France
†CNES, 31401 Toulouse, France

**Abstract.** The Sino-French SVOM mission (Space-based multi-band astronomical Variable Objects Monitor) has been designed to detect all known types of gamma-ray bursts (GRBs) and to provide fast and reliable GRB positions. In this study we present the SVOM pointing strategy which should ensure the largest number of localized bursts allowing a redshift measurement. The redshift measurement can only be performed by large telescopes located on Earth. The best scientific return will be achieved if we are able to combine constraints from both space segment (platform and payload) and ground telescopes (visibility).

The Sino-French SVOM mission (Space-based multi-band astronomical Variable Objects Monitor) has been designed to detect all known types of gamma-ray bursts (GRBs) and to provide fast and reliable GRB positions [1]. The goal of this study is to define the best SVOM pointing strategy which ensures the largest number of localized bursts allowing a redshift measurement. The redshift can only be measured by large telescopes located on Earth (8 m class). Therefore, the best scientific return will be achieved if one is able to combine constraints from both space segment (platform and payload) and ground large telescopes (visibility). Taking into account the scientific objectives and the orbital parameters, one should find (on a one year basis) a pointing law which fulfils the instrumental technical constraints as for instance power and thermal issues.

## SPACE SEGMENT REQUIREMENTS

The selected SVOM orbit is circular with an altitude of 600 km and an inclination angle of 30°. This orbit has a precession period of 60 days.

The SVOM payload combines wide field high energy instruments for GRB detection with narrow field telescopes sensitive in X-ray and visible wavelength for afterglow observations and position refinement [1, 2]. The accommodation of the SVOM multi-wavelength instrument set, aboard the same platform, has to face requirements which are different for each instrument. For example, the Sun, the Earth, and the Moon, have all different impacts, ranging from sensitivity losses at least, to destructive damages at most.

- ECLAIRs/CXG (Camera for X- and Gamma-rays). The CXG is 2D-coded mask imager sensitive from 4 to 250 keV. Its localization accuracy is better than 10 arcmin. The CXG field of view (FOV) is a double dihedral of $89° \times 89°$. The Sco X-1 X-ray source has to be kept outside the FOV in order to maintain the best sensitivity and to avoid false triggering. Similarly, the Galactic plane has to be avoided in order to prevent false triggering due to the bright X-ray binaries. The Earth and Sun transits within the FOV are non destructive.

- ECLAIRs/SXT (Soft X-ray Telescope). The SXT is a mirror focusing X-Ray telescope operating from 0.3 to 2 keV. Its FOV is 23 arcmin $\times$ 23 arcmin. Its localization accuracy is better than 10 arcsec. The Earth and Sun transits within the FOV are not destructive.
- GRM (Gamma-Ray Monitor). The GRM is a set of two gamma-ray detectors sensitive in the range 50 keV to 5 MeV. The ECLAIRs/CXG set of constraints is applicable to the GRM.
- VT (Visible Telescope). The VT is a 45 cm visible telescope operating from 400 to 950 nm. Its FOV is 21 arcmin $\times$ 21 arcmin. The Galactic plane should be avoided in order to avoid the optical wavelength absorption along the light of sight. The Earth transit within the FOV is non destructive. The Sun has to be maintained $90°$ away from the optical axis.

For the payload thermal issue, one should maintain one face of the satellite as cold as possible. Of course on the SVOM orbit, due to the presence of the earth, it is impossible to reach a pure cold face. Nevertheless the Sun has to be maintained more than $90°$ of the chosen cold face.

## GROUND BASED TELESCOPES REQUIREMENTS

The need of a fine spectroscopy (done by large visible/IR telescopes) imposes the GRB to be visible from places on ground where these telescopes are located[3]. Note that the most powerful instruments on ground are located close to the tropics. In order to be visible by a ground telescope, the sky zone observed by SVOM must respect the following condition: the optical axis (center of the FOV) has to be $40°$ above the local horizon. Three sites are considered : Maunea Kea, Cerro Paranal and Roque de los Muchachos. In addition, as for the payload visible instrument, the Galactic plane should be avoided because of the optical wavelength absorption along the light of sight.

## SUGGESTED ATTITUDE LAW

Taking into account the different constraints, one proposes the following scenario : The $X_{SVOM}$ axis aspect is optimized in order to avoid the presence of Sco X-1 and the Galactic plane within the ECLAIRs FOV. The $Y_{SVOM}$ axis is always maintained perpendicular to the Sun in order to allow a free rotation of the solar panels. For thermal reasons, the Sun is strictly kept inside the - $X_{SVOM}$/$Z_{SVOM}$ half plane.

The global scenario is a succession of anti-solar periods interleaved with avoidance periods during which the satellite passes away from the trigger "troublemakers" i.e. Sco X-1 and the Galactic plane. In some sky zones, the inertial pointing is kept during several

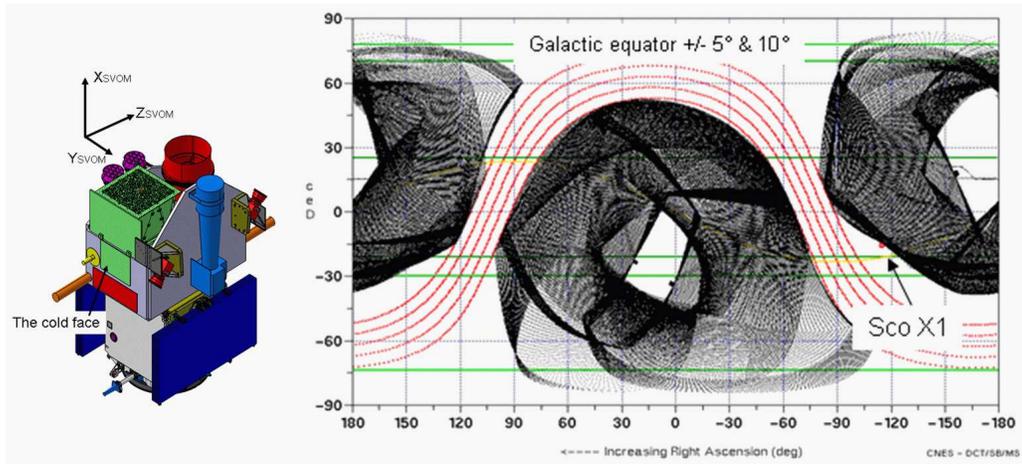

**FIGURE 1.** This figure presents on the left the reference axes and on the right the CXG field of view all along the year when the satellite is pointing the suggested attitude law.

days. The figure 1 shows reference axes and the ECLAIRs FOV track on the celestial sphere.

Let's remind that the sun orientation has to be kept between 180° and 90° from the $X_{SVOM}$ axis. Considering the CXG FOV and the pointing attitude constraints, it has been noticed that a value of 135° is an optimum regarding the slew maneuver in order to point a GRB after its detection. Note that in some cases the direction of the detected GRB could be close to a solar time of 6h or 18h.

In summary, the constraints lead to a non purely anti-solar pointing. Most of the year the $X_{SVOM}$ axis will be pointed at about 45° from the anti-solar direction, generally with the main component of the bias eastwards or westwards. The figure 2 presents the CXG exposure map all along the year when the satellite is pointing the suggested Attitude Law. Note that some sky zones are deeply observed.

## ANALYSIS OF GROUND TELESCOPES ACCES TO GRB

The visibility of the GRBs by the ground telescopes has been analyzed for the three selected sites. We have simulated 94 GRBs over one year. All these GRBs have been detected and pointed by the satellite for VT and ECLAIRs/SXT observations. The visibility duration is the total duration of visibility for one night, assuming the minimum site (40°) and a local sun elevation below -18°. These durations do not depend on the telescope longitude. Because of the non anti-solar pointing with the proposed offset of 45° for the $X_{SVOM}$ axis, some bursts close to solar times 6h or 18h can be detected by the ECLAIRs/CXG. For these bursts the ground telescopes observation can be difficult or even impossible. In our simulation, only 2 bursts are not visible by one of the 3 sites because of their solar time. The visibility durations are in the range between 0 and 8 hours. Among the 94 bursts detected and pointed by the satellite: 1 GRB is not visible on ground because its declination (79°) is out of the accessible range, 20 GRB are visible by 1 ground site, 25 GRB are visible by 2 ground sites, and 48 GRB are visible by the 3

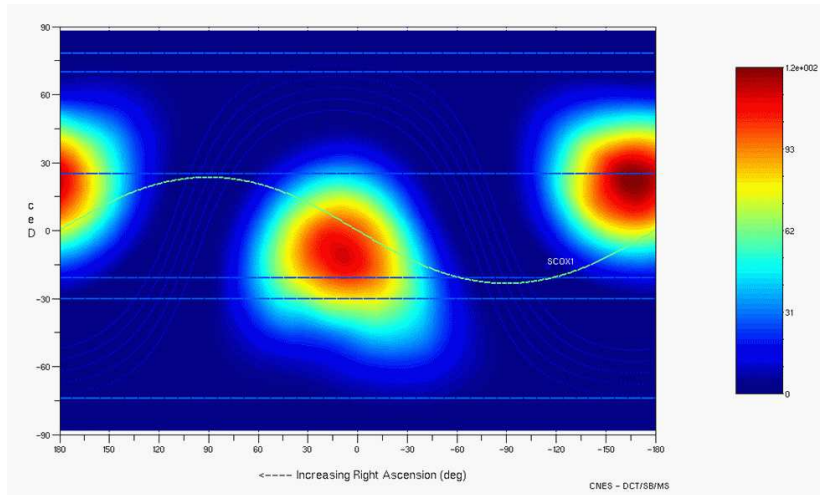

**FIGURE 2.** ECLAIRs/CXG exposure map all along the year when the satellite is pointing the suggested attitude law.

ground sites.

When there is no visibility on one site, this is because of the GRB declination which is out of the accessible range, except in the above mentioned two cases with solar time close to 6 hours. So in the great majority of cases, the GRB observation is possible by at least one of the three ground telescopes sites. The relatively large range of GRB directions solar times causes a reduction of the observations. In mean, the maximal observation duration (among the 3 sites) for one GRB is 6 hours.

We have computed the delay between the GRB detection and the possible observation by the ground telescopes. Considering the same 94 GRBs, we found that most of them are visible by a ground telescope at the time of their detection: among the 94 bursts, 69 GRB are immediately visible by one of the three ground telescope sites and 25 GRB are not visible immediately, but between 0 and 12 hours later.

## CONCLUSIONS

Taking into account the SVOM mission constraints, we propose a non purely anti-solar attitude law (the satellite will point at about $45°$ from the anti-solar direction) interleaved with avoidance periods during which the satellite passes away from the Sco X-1 source and the Galactic plane. The analysis of the ground telescope access to the SVOM GRB indicates that the suggested pointing strategy should ensure a large number of redshift measurements.

## REFERENCES

1. J. Paul, J. Wei, S. Zhang, and S. Basa, *these proceedings* (2008).
2. S. Schanne, *these proceedings* (2008).
3. S. Basa, *these proceedings* (2008).